# Copper-oxide Nanowires based Humidity Sensor

Ankit Vora[#1], Arvind K. Srivastava*

[#] *Shri Vaishnav Institute of Technology & Science, Indore, Madhya Pradesh, India*
* *Raja Ramanna Centre for Advanced Technology, Madhya Pradesh, India*
[1] ankit_vora@ieee.org

*Abstract* — This paper presents investigated results of copper-oxide nanowires used as a humidity sensor. Copper-oxide nanowires films were grown over cross-comb type gold electrodes on a $SiO_2$ substrate using thermal annealing technique, and its humidity sensitive characteristics were investigated through resistance across the gold electrodes. These copper-oxide nanowires films revealed high sensitivity and long-term stability with fast response time. It was found that resistance across gold electrodes of the fabricated sensor decreases with increase in humidity almost linearly on a logarithmic scale. It appears that copper-oxide nanowires can be used as low-cost humidity sensor with high output reliability and reproduction rate. The observations were carried out at room temperature (RT) and relative humidity (RH) in the range of 6% to 97%.

## I. INTRODUCTION

The semiconductor nanowires and nanorods have proved to have surprising new applications in sensors technology. Due to large surface-to-volume ratio, nanowires and nanorods find numerous possible applications in sensors, actuators, photonics landscape, charge transport, etc. They can be fabricated by many different processes depending upon the degree of quality in terms of orientation, length, width, shape, and identicalness of nanowires/ nanorods. They can be easily produced in bulk at a low-cost with optimized ordered length, width, and orientation by many processes. Thermal annealing is one of the best techniques as far as cost, and optimized quality is concerned. Thermal annealing of copper foil in air enables fabrication of bulk, low cost, nanowires with length as high as 8-10μm and width as low as 20-30nm with high ordered array of CuO nanowires as reported in [1]. Nanowires of semiconducting materials used as gas sensors (or humidity sensors) have been reported in [2], [3] and [4], but most of them show poor repeatability when used as a practical humidity sensor. The most significant advantage of copper-oxide nanowires is the ease of fabrication and calibration, with low fabrication cost. Due to small band-gap (~1.2eV), it shows better humidity sensitive characteristics (resistance) as compared to other gas sensors and its humidity sensitive characteristics (resistance) can be approximated as almost linearly depreciating (logarithmic scale) with an increase in relative humidity (RH). Gas sensitive characteristic of copper-oxide nanowire film is very large as compared to other types of semiconducting nanowires because of high length to width ratio and high porosity of CuO nanowires film. A Dense forest of copper-oxide nanowires can be easily fabricated on $SiO_2$ substrate by simple thermal annealing of copper foil on required substrate making it a low-cost sensor. It is experimentally demonstrated that CuO nanowires film show a promising application for humidity sensors. ZnO nanowire film based humidity sensor has been reported in [11], but ZnO has a wide bandgap of 3.4eV which is the cause of poor conductivity of ZnO nanowires, and it is also a piezoelectric material, which can show various unpredictable observations due to the piezoelectric effect. Therefore, to overcome this problem, narrow band gap semiconducting material was required having band gap around 1-1.5eV, and stability was an important criterion. Hence, copper-oxide nano-wires were chosen to experiment.

## II. EXPERIMENT

As the first step, the $SiO_2$ substrate was cleaned in an ultrasonic bath in acetone for about 30 minutes. Then gold was deposited on $SiO_2$ substrate in the cross-comb form to be used as electrodes. A thin copper foil, almost same in size as that of the substrate was cleaned in dilute HCl and was put over the substrate, and the entire substrate was kept in a furnace at a temperature of 500°C for thermal annealing in oxygen for 4 hours. After that, the substrate was allowed to cool down till room temperature was achieved. High-quality CuO nanowires were grown on $SiO_2$ substrate as shown in SEM micrograph of CuO nanowires in Fig. 1. TEM and XRD analysis of CuO nanowire is also shown in Fig. 3 and 4 respectively. The growth of CuO nanowires was of high yield and ordered in nature as it can be seen in Fig. 1, nanowires with length as high as 6-10μm and width as low as 20-40nm were observed which are of very high quality as far as humidity sensing application is concerned.

### A. Experimental setup

The entire experiment was carried inside a sealed glass chamber. An inlet for water vapor was provided for humidity control, and sensor (substrate) was kept over Thermo-Electric-Cooler (TEC) for temperature control of sensor, for maintaining the temperature at 25°C (RT) during the whole experiment. The electrical properties of the materials were investigated using an LF impedance analyzer and electrometer. The schematic of the sensor is shown in Fig. 2 which is almost similar as used in [11]. The controlled humidity environments were achieved using anhydrous $P_2O_5$, silica and saturated aqueous solutions of various salts at an ambient temperature of 25°C, which yielded fixed relative humidity for calibration purpose. These RH levels were independently monitored by using a standard hygrometer kept inside the sealed glass chamber. The details of salts used and there yielded RH is given in table 1. The humidity level was gradually increased, and the corresponding resistance across the sensor electrodes was measured, RH level was varied from 6% to 97%.

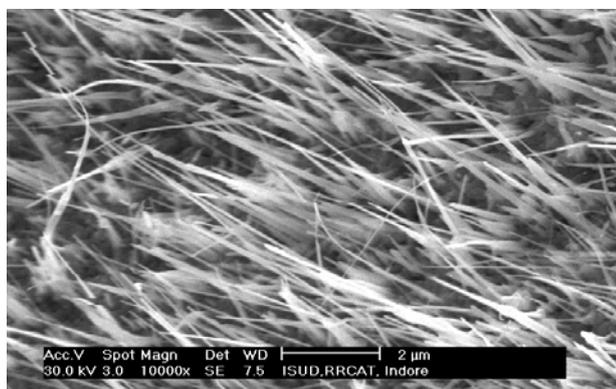

Fig. 1 SEM image of the CuO nanowires film grown over $SiO_2$ substrate by thermal annealing method.

### III. RESULTS AND DISCUSSIONS

The desirable characteristics of good humidity sensors are high sensitivity, chemical and thermal stability, reproducibility, low operation temperature, low cost and long life. So far, there is no material known to fulfill all those requirements simultaneously [5], [6] and [7]. CuO nanowires were used for humidity sensitive characteristics against its resistance. Gold electrodes were chosen for being inert during thermal annealing of copper foil to prevent electrodes from oxidation. TEM analysis of a single nanowire revealed many details like roughness and dimensions of CuO nanowire and XRD analysis showed crystal structure to be monoclinic crystal system and its lattice parameters being a = 4.6837Å, b = 3.4226Å, c = 5.1288Å, α = 90°, β = 99.54°, γ = 90°. Thermal annealing of Cu foil at 500°C results in the growth of dense forest of CuO nanowires film spread almost uniformly over the whole substrate with length as high as 6-10μm and width as low as 20-40nm with high porosity. The growth process of the CuO nanowires can be interpreted by means of the vapor–liquid–solid mechanism [8], [9].

TABLE I
SALTS USED FOR CALIBRATION OF RH AND CONTROLLING HUMIDITY CONDITIONS

| Salts used in Experiment | Solubility (g/100mL) at 25°C | RH (%) |
|---|---|---|
| Lithium Bromide | 161 | 6.37 |
| Lithium Chloride | 83.2 | 11.3 |
| Potassium Acetate | 257 | 22.51 |
| Magnesium Chloride | 167 | 32.8 |
| Potassium Carbonate | 114 | 43.16 |
| Magnesium Nitrate | 125 | 52.89 |
| Sodium Bromide | 75 | 57.57 |
| Potassium Iodide | 142 | 68.86 |
| Sodium Chloride | 36 | 75.3 |
| Potassium Chloride | 36 | 84.34 |
| Potassium Nitrate | 36 | 95 |
| Potassium Sulphate | 12 | 97.3 |

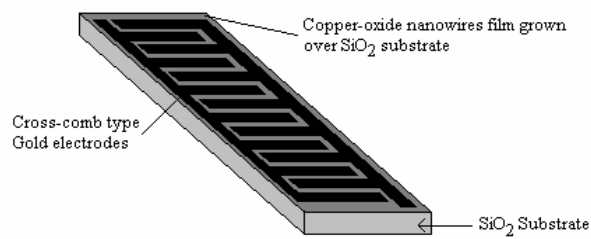

Fig. 2. Schematic diagram of CuO nanowires film based humidity sensor.

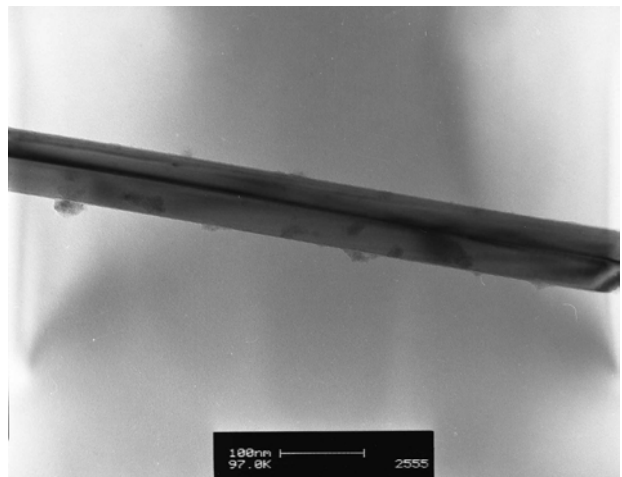

Fig. 3. TEM analysis of single nanowire, scale equal to 100nm

Investigating the resistive-type sensor's I–V characteristics measured in the RH range of 6-97% indicated a strong dependence of electrical resistance on RH. The results of electrical resistance measurement as a function of RH at a fixed ambient temperature of 25°C are depicted in Fig. 5. It can be seen that the resistance of the CuO nanowires film decreases with increasing relative humidity. The resistance of CuO nanowires sample changes linearly (logarithmic scale) with RH for approximately 3 ($10^8$ -$10^5$ Ω) orders of magnitude on the semi-log scale over the range of 6–97% RH, showing high sensitivity and excellent linearity.

The resistance variations with time for CuO nanowires humidity sensor was also carried out. The measurements were repeated at 25°C, every day for one month. Slight variation in resistance was observed most of the time, but resistance variation was observed to be less than 1%, at each humidity region, for one month. Hence it could be said that resistance of CuO nanowires film is quite stable to the exposure to water in the air. But a little problem of hysteresis was observed in the sensor; this might be due to the trapped moisture inside the pores of nanowires film. Although the change was minor, it was frequently observed. Therefore, TEC was used first to heat the sensor and then cool down slowly; this resulted in sweeping out the trapped moisture, and hence again output was nearly same as earlier.

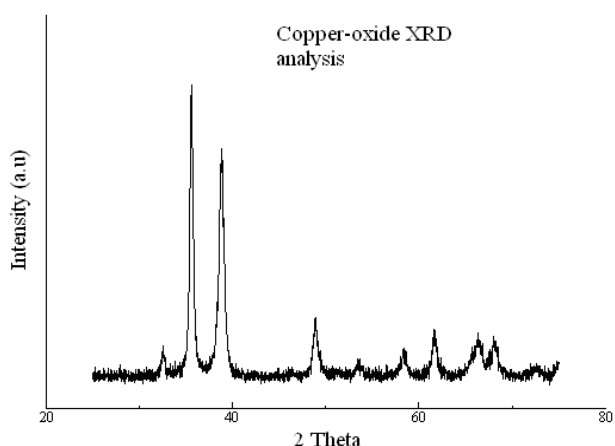

Fig. 4. XRD analysis of CuO nanowire film

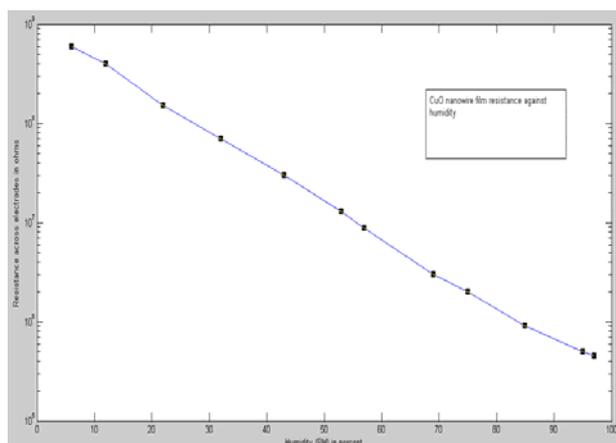

Fig. 5. Relative humidity vs. DC resistance plots at 25°C for CuO nanowires. Resistance is plotted against y-axis on a logarithmic scale, and the relative humidity is plotted against x-axis.

Here, a proposed mechanism for humidity sensing characteristics of CuO nanowires is provided on the similar basis of proposed mechanism of ZnO nanowires gas sensing characteristics as described in [11]. The substantial increase in conductivity with increased RH of CuO nanowires is also related to the adsorption of water molecules. There is a lot of space between two adjacent CuO nanowires as their growth is irregular, and the films consisting of CuO nanowires are similar to porous structure materials, having a higher specific surface area. Hence this kind of nanowires films absorbs moisture readily. At low humidity, tips and defects of the nanorods and nanowires present a high local charge density and a strong electrostatic field, which promotes water dissociation [10]. The dissociation provides protons as charge carriers for hopping transport. At high humidity, liquid water condenses on the nanorods and nanowires, and electrolytic conduction between nanorods and nanowires takes place along with protonic transport. This suggested mechanism can explain the decrease in sensor impedance with increasing humidity.

## IV. CONCLUSION

To conclude, CuO nanowires were fabricated on silicon dioxide substrate using thermal annealing method and very high quality ordered nanowires film was synthesized on $SiO_2$ substrate; cross-comb shape gold electrodes were deposited before nanowires synthesis. The nanowires film was realized to work as an effective humidity sensor, and electrical resistance was found to be linearly varying on a logarithmic scale with the increase in humidity from 6% to 97% of RH level. This nanowires film shows high humidity sensing characteristics, and it was found to have a high degree of reproducibility and stability. Therefore, copper-oxide nanowires can be used as an effective humidity sensor with low manufacturing cost.

## ACKNOWLEDGMENT

The authors want to thank Mahendra Babu for discussion, TEM, and SEM analysis.